\documentclass[a4paper,11pt]{article}
\pdfoutput=1

\usepackage{jcappub}

\usepackage[T1]{fontenc}
\usepackage{multirow}
\usepackage{bm}
\usepackage{mathtools}
\usepackage[version=4]{mhchem}
\usepackage{siunitx}
\usepackage[normalem]{ulem}
\usepackage{comment}
\usepackage[export]{adjustbox}
\usepackage{xcolor}

\newcommand{\be}{\begin{equation}}
\newcommand{\ee}{\end{equation}}
\newcommand{\de}{{\rm d}}
\newcommand{\bs}[1]{\textbf{\textsf{#1}}}

\compress
\title{Stairway to Axions: the cross-correlation of birefringence and galaxies from NPIPE and Quaia}

\author[a,b]{S.~Arcari,}
\author[c,d,e]{N.~Bartolo,}
\author[f,g]{G.~Fabbian,}
\author[h]{A.~Greco,}
\author[i,j,a]{A.~Gruppuso,}
\author[a,b]{M.~Lattanzi,}
\author[a,b]{P.~Natoli,}
\author[a,b]{L.~Pagano,}
\author[a,b]{and G.~Zagatti}

\affiliation[a]{Dipartimento di Fisica e Scienze della Terra, Università degli Studi di Ferrara, Via G. Saragat 1, I-44122 Ferrara, Italy}
\affiliation[b]{Istituto Nazionale di Fisica Nucleare, Sezione di Ferrara, Via G. Saragat 1, I-44122 Ferrara, Italy}
\affiliation[c]{Dipartimento di Fisica e Astronomia “Galileo Galilei”, Universià degli Studi di Padova, Via Marzolo 8, I-35131 Padova, Italy}
\affiliation[d]{Istituto Nazionale di Fisica Nucleare, Sezione di Padova, Via Marzolo 8, I-35131 Padova, Italy}
\affiliation[e]{Istituto Nazionale di Astrofisica - Osservatorio Astronomico di Padova, Vicolo dell’Osservatorio 5, I-35122 Padova, Italy}
\affiliation[f]{Kavli Institute for Cosmology Cambridge, Madingley Road, Cambridge CB3 0HA, UK}
\affiliation[g]{Institute of Astronomy, Madingley Road, Cambridge CB3 0HA, UK}
\affiliation[h]{Department of Astronomy, University of Florida, 211 Bryant Space Science Center, Gainesville, FL 32611, USA}
\affiliation[i]{Istituto Nazionale di Astrofisica - Osservatorio di Astrofisica e Scienza dello Spazio di Bologna, Via Gobetti 101, I-40129 Bologna, Italy}
\affiliation[j]{Istituto Nazionale di Fisica Nucleare, Sezione di Bologna, Viale Berti Pichat 6/2, I-40127 Bologna, Italy}

\emailAdd{stefano.arcari@unife.it}
\emailAdd{nicola.bartolo@pd.infn.it}
\emailAdd{fabbiang@cardiff.ac.uk}
\emailAdd{alessandro.greco@ufl.edu}
\emailAdd{alessandro.gruppuso@inaf.it}
\emailAdd{massimiliano.lattanzi@fe.infn.it}
\emailAdd{paolo.natoli@unife.it}
\emailAdd{luca.pagano@unife.it}
\emailAdd{giorgia.zagatti@unife.it}

\abstract{We present the first measurement of the cross-correlation between anisotropic birefringence and galaxy number counts, utilizing polarization data from {\it Planck} NPIPE and the Quaia quasar catalog. By employing a QML/pseudo-$C_\ell$ combined estimator, we compute the angular power spectrum up to $\ell=191$ from birefringence and clustering maps at $N_{\rm side}=64$. Our analysis indicates that the observed spectrum is well consistent with the null-hypothesis, with a probability to exceed of 37\% and an estimated scale-invariant amplitude of $A^{\mathcal{D}_\ell}=(2.22\pm2.09)\times10^{-4}\,\deg$, at the 68\% confidence level. Finally, we derive constraints on the axion-parameters within an early dark energy model of birefringence. Our findings reveal an unprecedented upper bound on the axion-photon coupling down to $g_{\phi\gamma}=10^{-15}\;\si{\giga\electronvolt^{-1}}$ for masses around $10^{-32}\,\si{\electronvolt}$ and high initial misalignment angles. This result opens a previously unexplored window in parameter space, providing the first constraint in this ultra-light mass regime.}

\begin{document}
\maketitle

\section{Introduction}
\label{sec:intro}
The Cosmic Microwave Background (CMB) is a pivotal tool for exploring the early universe and fundamental cosmology. Initial CMB experiments \citep{Mather:1990tfx,Jones:2005yb,WMAP:2012nax}, including the {\it Planck} satellite \citep{Planck:2013pxb,Planck:2015fie,Planck:2018vyg, Planck:2018nkj}, focused on temperature anisotropies, confirming the standard cosmological model while revealing subtle anomalies. Recently, research has shifted towards CMB polarization, which provides deeper insights into cosmic inflation, parity violation, and the early universe's dynamics. By analyzing these polarization patterns, researchers can place constraints on new physics parameters, providing insights into phenomena beyond the standard cosmological model, including dark matter, dark energy, and other exotic fields.

A notable effect in CMB studies is cosmic birefringence (CB) \citep{Carroll:1989vb, Carroll:1998zi, Harari:1992ea,Li:2008tma, Pospelov:2008gg, Finelli:2008jv,Balaji:2003sw, Liu:2016dcg, Capparelli:2019rtn, Caldwell:2011pu, Galaverni:2014gca, Komatsu:2022nvu,Myers:2003fd, Gubitosi:2009eu, Gubitosi:2010dj,Kostelecky:2007zz,Caloni:2022kwp}, the rotation of the linear polarization plane of CMB radiation. A well-known modification of the standard model involves adding a Chern-Simons term to the Maxwell Lagrangian density, which describes an interaction between photons and a new scalar field $\phi$ \citep{Carroll:1989vb, Carroll:1998zi, Harari:1992ea,Li:2008tma, Pospelov:2008gg, Finelli:2008jv}:
\begin{equation}
\mathcal{L} \supset -\frac{1}{2}\,g^{\mu\nu}\partial_\mu\phi\,\partial_\nu\phi - V(\phi) -\frac{1}{4}\,g_{\phi\gamma}\,\phi\,F_{\mu\nu}\tilde{F}^{\mu\nu} \;,
\label{eq:lagrangian}
\end{equation}
where \( g^{\mu\nu} \) is the metric tensor, \( V(\phi) \) the field potential, \( g_{\phi\gamma} \) the field-to-photon coupling, and \( \tilde{F}^{\mu\nu} = \epsilon^{\mu\nu\rho\sigma}F_{\rho\sigma}/2 \) the Hodge dual of the Maxwell tensor \( F_{\mu\nu} \). The last term of eq.~\eqref{eq:lagrangian} induces parity violation when the field $\phi$ retains spacetime dependence, and appears naturally for axion-like particles (ALPs) \citep{Witten:1984dg,Conlon:2006tq,Svrcek:2006yi,Vilenkin:1982ks,Huang:1985tt,Davis:1986xc,Arvanitaki:2009fg, Marsh:2015xka, Kitajima:2022jzz, Jain:2021shf,Abbott:1982af,Lin:2022niw,Liu:2016dcg,Fujita:2020ecn,Caldwell:2011pu,Preskill:1982cy,Dine:1982ah,Nakagawa:2021nme,Obata:2021nql,Hlozek:2014lca,Poulin:2018dzj,Kim:2021eye,Greco:2022xwj, Gonzalez:2022mcx, Gasparotto:2023psh,Greco:2024oie,Poulin:2018cxd,Capparelli:2019rtn,Choi:2021aze,Murai:2022zur,Gasparotto:2022uqo,Kamionkowski:2022pkx,Arcari:2024nhw}. The primary effect on CMB is a mixing between E- and B-modes, leading to non vanishing TB and EB cross-correlations. Recent works \citep{Namikawa:2021gbr, Minami:2020odp, Komatsu:2022nvu, Fujita:2022qlk, Diego-Palazuelos:2022dsq, Eskilt:2022wav} have exploited novel analysis techniques yielding a hint of detection (at $3\sigma$) for the isotropic CB angle. Moreover, one expects anisotropies in the birefringence angle to arise naturally from the fluctuations of the scalar. These variations across the sky have been studied extensively through CMB data \citep{Gruppuso:2020kfy, Bortolami:2022whx, BICEPKeck:2022kci,SPT:2020cxx,Zagatti:2024jxm} and proposed as a tool to constraint early dark energy (EDE) models and axions \citep{Caldwell:2011pu, Capparelli:2019rtn, Greco:2022ufo,Greco:2022xwj,Arcari:2024nhw}.

In this work, we show, for the first time, the cross-correlation between anisotropic birefringence and the spatial distribution of galaxies, as measured by the combination of CMB polarization data from {\it Planck} PR4 \citep{Planck:2020olo,Tristram:2023haj} and the all-sky quasar catalog Quaia \citep{Storey-Fisher:2023gca,Alonso:2023guh,Piccirilli:2024xgo,Gaia:2023}. We exploit a pseudo-$C_\ell$ approach (supported by QML estimates at low-$\ell$'s) over maps for anisotropic birefringence derived through the EB-estimator of ref. \citep{Zagatti:2024jxm}, based on \citep{Gluscevic:2012me,Namikawa:2020ffr}, and for galaxy number counts derived from Quaia \citep{Storey-Fisher:2023gca}. The associated covariance is consistently calculated using 400 polarization plus noise simulations from {\it Planck} NPIPE \citep{Planck:2020olo}, alongside an equal number of realizations from the quasar catalog. Our results reveal a cross-correlation well consistent with the null-hypothesis across the whole range of multipoles. Finally, we exploit this novel signal to constraint axion-parameters through a Gaussian likelihood approach over theoretical spectra derived with a properly modified version of the Boltzmann code {\tt CLASS} \citep{DiDio:2013bqa,Lesgourgues:2011re}. We introduce an unprecedented upper bound on the axion-photon coupling $g_{\phi\gamma}$ within the ultralight mass range $m_\phi=\left[10^{-33},\;10^{-28}\right]\,\si{\electronvolt}$. In conclusion, we expand on the theoretical introduction of the birefringence-LSS cross-correlation of ref. \citep{Arcari:2024nhw} by measuring it for the first time, and show its exceptional potential in constraining axion-like physics.

The paper is organized as follows: in section \ref{sec:data} we describe the data and simulation sets used to derive the birefringence-galaxy cross-correlation and related covariance, whilst in section \ref{sec:meth} we briefly summarize its theoretical interpretation and the adopted estimation techniques. In section \ref{sec:res} we show the measured cross-correlation, alongside our constraints on the axion-photon coupling. Finally conclusions are drawn in section \ref{sec:conc}. Appendix \ref{app:A} serves as an informative summary of possible implications on the results of changing the multipole-binning scheme.

Throughout the paper we assume a spatially flat $\Lambda$CDM cosmology with cosmological parameters as derived by the final {\it Planck} data release \citep{Tristram:2023haj}.\footnote{Let us remind that ALPs are an extension of the underlying cosmological model.}

\section{Datasets}
\label{sec:data}
This section serves as a brief description of the data products employed in our study. Specifically, we review the procedure to obtain anisotropic birefringence maps from {\it Planck} NPIPE \citep{Planck:2020olo} and galaxy overdensity maps from Quaia \citep{Storey-Fisher:2023gca}.

\subsection{\emph{Planck} NPIPE}
\label{subsec:data_alpha}
\begin{figure}[t]
\centering
\includegraphics[width=0.95\hsize]{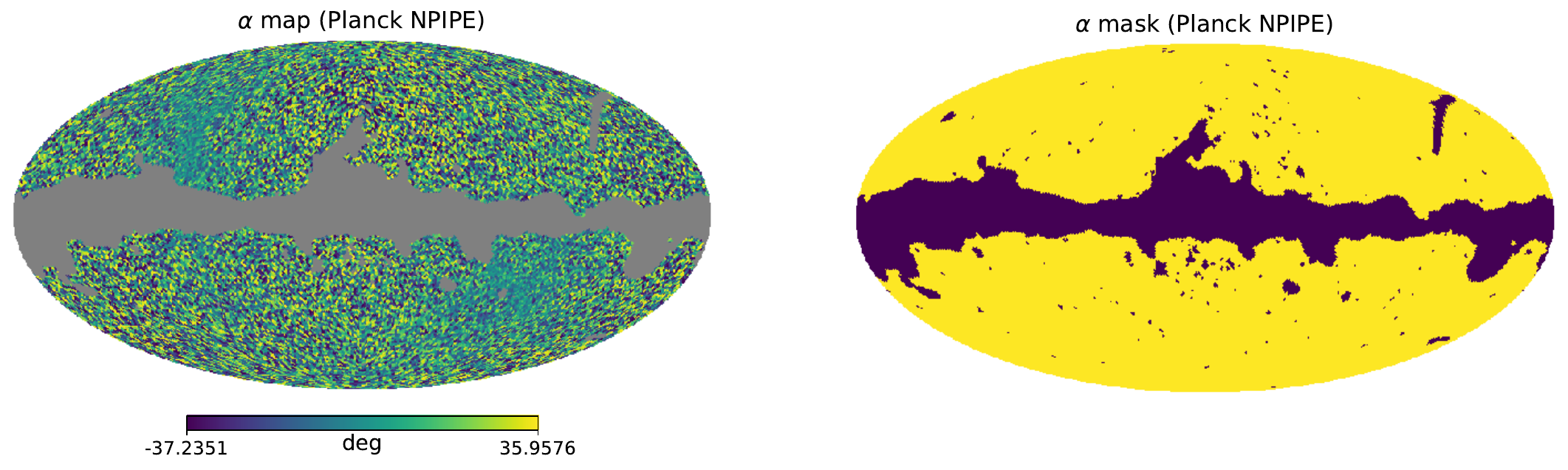}
\caption{Anisotropic birefringence map derived from {\it Planck} NPIPE polarization data with the EB-estimator of ref.~\citep{Zagatti:2024jxm} (\emph{Left}), and corresponding sky mask (\emph{Right}).}
\label{fig:data_alpha}
\end{figure}
Cosmic birefringence induces a rotation of linearly polarized light by an angle $\alpha({\bf \hat n}) = \bar\alpha+\delta\alpha({\bf \hat n})$, where the isotropic contribution $\bar\alpha$ arises from the background field, while field perturbations induce a dependence on the direction on the sky ${\bf \hat n}$ in the anisotropic counterpart $\delta\alpha({\bf \hat n})$. As a consequence, the observed CMB E- and B-modes are rotated and mixed \citep{Lue:1998mq,Kamionkowski:2008fp} and inherit information on the rotation field. Harmonic estimators can be exploited to reconstruct birefringence from CMB polarization \citep{Gluscevic:2012me,Namikawa:2020ffr}. In particular, focusing on the EB cross-correlation only allows to avoid the cosmic variance associated with the temperature field and rises as a promising tool, given the higher predicted signal-to-noise ratio with respect to e.g. TB. Following the EB-estimator implemented by \citep{Zagatti:2024jxm}, we are able to reconstruct the harmonic coefficients associated to anisotropic birefringence from {\it Planck} NPIPE data. The related {\tt HEALPix} \citep{Gorski:2004by} map at $N_{\rm side}=64$ is shown in the left panel of fig.~\ref{fig:data_alpha}. The latter is masked according to {\it Planck} NPIPE fiducial, with a corresponding sky fraction $f_{\rm sky}=78\%$ (see right panel of fig.~\ref{fig:data_alpha}). In addition, this dataset provides 400 simulations of polarization$\,+\,$noise that we use to obtain realizations of the anisotropic birefringence sky for covariance estimation, as discussed in section~\ref{subsec:cov}.

The birefringence maps used are generated with a minimum CMB multipole $\ell_{\rm min}^{\rm CMB} = 2$, in order to account for the polarization generated at all redshifts, particularly from both the recombination and reionization epochs. Ref.~\citep{Zagatti:2024jxm} demonstrates that the choice of $\ell_{\rm min}^{\rm CMB}$ has negligible impact on the birefringence auto-correlation, and we have tested that the same is true for the underlying cross-correlation. In this study, we aim to perform a ‘blind’ analysis to extract information on axion parameters from the complete birefringence signal, without isolating contributions from different cosmic epochs. Thus, the estimator from ref.~\citep{Zagatti:2024jxm} remains unaltered, though a different approach would be needed if our goal were to analyze birefringence at a specific redshift, as suggested by \citep{Namikawa:2024sax}. Given the extensive axion parameter space and the variability in dominant redshifts within it (see \citep{Arcari:2024nhw}), we avoid assumptions on the expected signal and proceed with an analysis on maps incorporating all birefringence contributions.

\subsection{Quaia}
\label{subsec:data_quaia}
\begin{figure}[t]
\centering
\includegraphics[width=0.95\hsize]{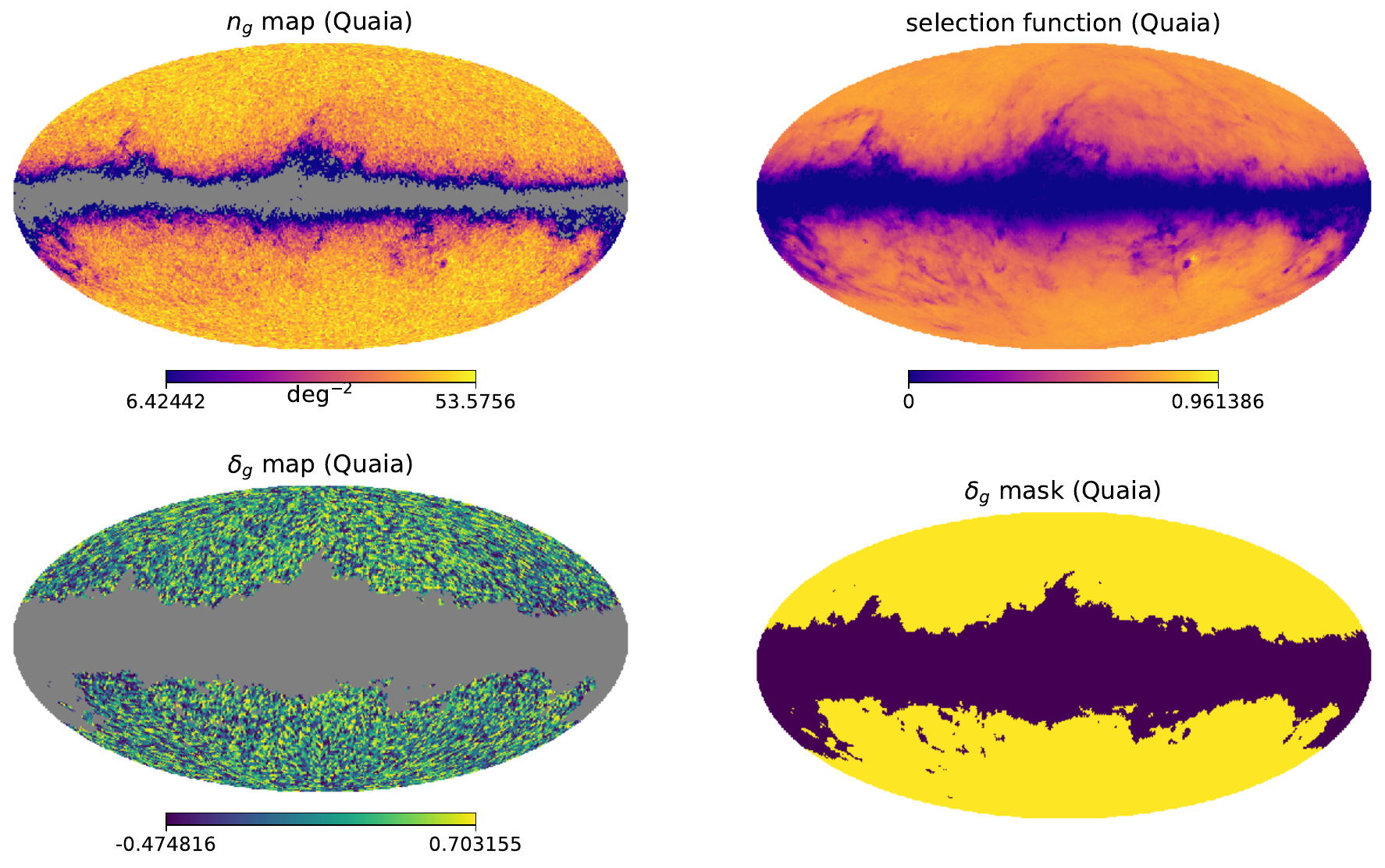}
\caption{(\emph{Top left}): pixel density of sources from the Quaia quasar catalog \citep{Storey-Fisher:2023gca}; the map includes almost 1.3 million objects. (\emph{Top right}): related selection function. (\emph{Bottom left}): galaxy overdensity from Quaia derived as outlined in section~\ref{subsec:data_quaia}. (\emph{Bottom right}): Quaia sky mask.}
\label{fig:data_quaia}
\end{figure}

Quasars offer as a powerful tool to probe accretion physics \citep{Sunyaev:1970bma,Yu:2020}, galaxy formation \citep{Kormendy:2013dxa}, supermassive black holes \citep{Hopkins:2005fb} and black hole evolution \citep{Powell:2020}, gravitational lensing \citep{Claeskens:2002}, halo masses \citep{DiPompeo:2017} and the intergalactic medium \citep{Rauch:1998xn}. Furthermore, quasars are exceptional tracers of large-scale-structure cosmology due to their position within peaks of the dark matter distribution, and can be exploited to constrain cosmological parameters \citep{Garcia-Garcia:2021unp,Alonso:2023guh,eBOSS:2020gbb,Leistedt:2014zqa,eBOSS:2019sma,SDSS:2004vxk,eBOSS:2017cqx,Sherwin:2012mr,Menard:2002vz,Risaliti:2015zla}. We consider the {\it Gaia-unWISE} Quasar catalog (Quaia), which is an all-sky sample comprising almost 1.3 million objects with magnitude $G<20.5$, up to redshift $z\sim3$. The catalog stems from the combination of quasar candidates in the third data release of {\it Gaia} \citep{Gaia:2023} and the {\it unWISE} \citep{Lang:2014,Meisner:2019lbf} infrared photometry, based on the Wide-field Infrared Survey Explorer ({\it WISE}$\;$) \citep{Wright:2010qw}. The corresponding {\tt HEALPix} distribution of sources\footnote{The Quaia related data products have been produced following ref.~\citep{Storey-Fisher:2023gca} and publicly available at \url{https://zenodo.org/records/8060755}.}, at $N_{\rm side}=64$, is illustrated in the top left panel of fig.~\ref{fig:data_quaia}. Following refs.~\citep{Storey-Fisher:2023gca,Alonso:2023guh,Piccirilli:2024xgo} we obtain the projected galaxy-overdensity in each pixel ${\bf \hat{n}}$ as $\delta_g({\bf\hat{n}}) = N({\bf\hat{n}})/(\bar{N}\omega({\bf\hat{n}}))-1$, where $N({\bf\hat{n}})$ is the number of quasars in the pixel and $\omega({\bf\hat{n}})$ the related selection function (shown in the top right panel of fig.~\ref{fig:data_quaia}). Finally, $\bar{N}=\sum_{\bf\hat{n}}N({\bf\hat{n}})/\sum_{\bf\hat{n}}\omega({\bf\hat{n}})$ is the mean number of sources per pixel. The resulting galaxy-overdensity map is depicted in the bottom left panel of fig.~\ref{fig:data_quaia}. We mask all pixels in the sky where the selection function is lower than 0.5 (see bottom right panel of fig.~\ref{fig:data_quaia}). To match the 400 birefringence simulated maps for covariance estimation, we generate an equal number of realizations of $\delta_g$ by reshuffling a random catalog with ten times the sources of the data catalog outlined above.\footnote{Each galaxy realization conserves the total number of sources of the data-map.}

\section{Methodology}
\label{sec:meth}
We review here the theoretical formulation of the birefringence-galaxy cross-correlation, first proposed in \citep{Arcari:2024nhw}, with particular attention to key ingredients related to the chosen data (see section~\ref{sec:data}). We then summarize the strategy adopted to extract the observed underlying cross-correlation.

\subsection{Theoretical model}
\label{subsec:theory}
\begin{figure}[t]
\centering
\includegraphics[width=0.55\hsize]{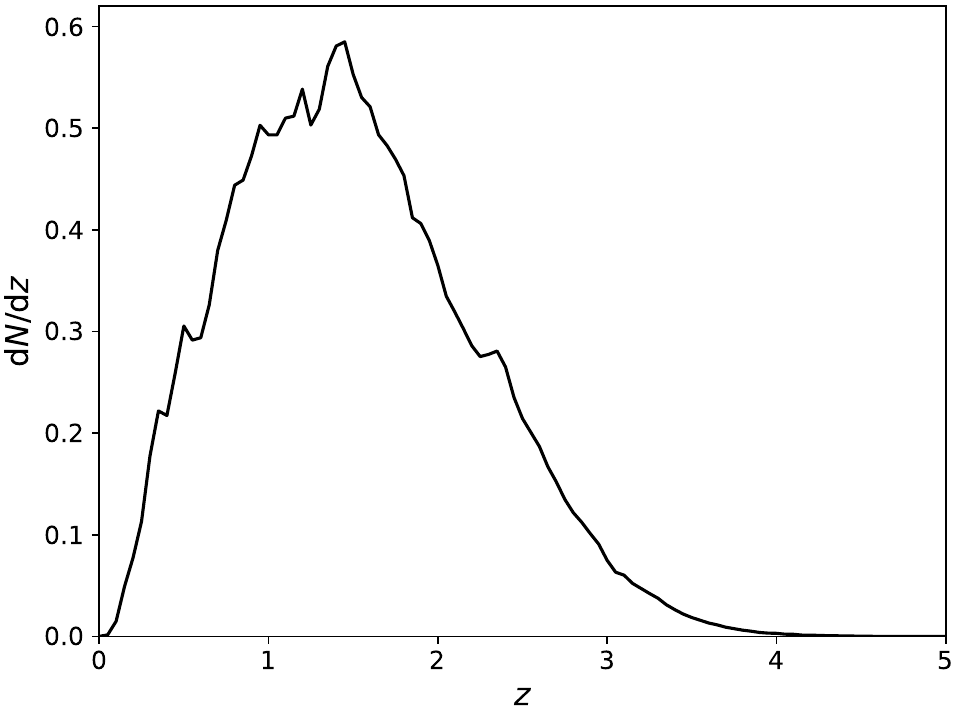}
\caption{Redshift distribution of sources within the Quaia catalog \citep{Storey-Fisher:2023gca}, used as the selection function in eq.~\eqref{eq:galkernel} for the computation of the theoretical cross-correlation.}
\label{fig:dndz}
\end{figure}
In this study, we regard two different probes: the anisotropic birefringence $\alpha({\bf\hat{n}})$ and the spatial distribution of galaxies as measured by the Quaia quasar overdensity $\delta_g({\bf\hat{n}})$. As discussed in section~\ref{sec:intro}, the former can be generated by parity violating couplings of ALPs to the electromagnetic sector, and we expect a non-zero correlation due to both probes being sourced by the metric perturbations and related gravitational potentials \citep{Greco:2022xwj,Greco:2022ufo}. This has been extensively studied theoretically in ref.~\citep{Arcari:2024nhw} and presented, for the first time, as a valuable tool to explore the axion-parameter space. The angular cross-correlation reads\footnote{Here, adiabatic initial conditions are assumed.}
\be\label{eq:alphaG}
    C_\ell^{\alpha g} = 4\pi\int \frac{{\rm d}k}{k}\;\mathcal{P}_\mathcal{R}(k)\Delta_\ell^\alpha(k)\Delta_\ell^{g}(k) \,,
\ee
where $\mathcal{P}_\mathcal{R}$ is the primordial power spectrum, while $\Delta_\ell^\alpha(k)$ and $\Delta_\ell^{g}(k)$ are the respective kernels for birefringence and galaxies \citep{DiDio:2013bqa,Challinor:2011bk,Bonvin:2011bg}:
\begin{align}
\label{eq:alphakernel}
     \Delta^\alpha_\ell(k) &= g_{\phi\gamma}\int_0^{\tau_0}{\rm d}\tau\;g(\tau)T_{\delta\phi}(\tau,\,k)j_\ell\left[k(\tau_0-\tau)\right] \,,\\
\label{eq:galkernel}
     \Delta^{g}_\ell &= \int{\rm d}z\;\frac{{\rm d}N}{{\rm d}z}b(z)S_D(z,\,k)j_\ell\left[k\,r(z)\right] \,.
\end{align}
In eq.~\eqref{eq:alphakernel}, $g(\tau)$ is the full photon visibility function with respect to conformal time $\tau$, $T_{\delta\phi}(\tau,\,k)$ is a transfer function between the primordial power spectrum $\mathcal{P}_\mathcal{R}(k)$ and that of the field fluctuations $\delta\phi$, and $j_\ell\left[k(\tau_0-\tau)\right]$ the $\ell$-th spherical Bessel function. This kernel is governed by the Klein-Gordon equation of axion-perturbations
\be\label{eq:EoMpert}
    \delta\phi''+2\mathcal{H}\,\delta\phi'+\left(k^2+a^2\frac{\de^2V}{\de\bar{\phi}^2}\right)\delta\phi=-\frac{1}{2}h'\bar{\phi}' \;,
\ee
with $\mathcal{H}$ being the conformal Hubble parameter, $a$ the scale factor, $h$ the metric perturbation in synchronous gauge, $\bar \phi$ the background field, and primes denoting derivatives with respect to conformal time $\tau$. For our analysis, we choose an axion-potential $V$ stemming from EDE models of the string axiverse \citep{Svrcek:2006yi,Arvanitaki:2009fg,Kamionkowski:2014zda}:
\be\label{eq:potential}
    V(\phi) = m_\phi^2\,f_a^2\left[1-\cos{\frac{\phi}{f_a}}\right]^3 \,.
\ee
Here, $f_a$ is the scale of the spontaneous breakdown of a continuous symmetry\footnote{We fix it to the {\it Planck} mass.} and $m_\phi$ the axion mass. In summary, the axion evolution and its impact on birefringence will be controlled by the axion-photon coupling $g_{\phi\gamma}$, the axion mass and its initial misalignment $\theta_i=\phi_i/f_a$ (setting the initial condition of the axion field as well\footnote{The considered potential arises non-perturbatively from phase-space considerations. Hence, the initial misalignment is bound to live in the $(-\pi,\,\pi]$ domain \citep{Marsh:2015xka,Conlon:2006tq}.}).

In eq.~\eqref{eq:galkernel}, ${\rm d}N(z)/{\rm d}z$ is the galaxy redshift distribution, $b(z)$ the linear bias between the galaxy and matter overdensity, $S_D$ a source function related to the growth of structure and the Bardeen potentials, and $r(z)$ the comoving radial distance. The former is taken directly from the data products of Quaia discussed in section~\ref{subsec:data_quaia} and shown in fig.~\ref{fig:dndz}. For the galaxy bias, we follow the prescription of \citep{eBOSS:2017ozs}:
\be\label{eq:bias}
    b(z) = 0.278\left[(1 + z)^2 - 6.565\right] + 2.393 \,.
\ee

Eq.~\eqref{eq:alphaG} and the aforementioned axion model and clustering ingredients have been implemented in a modified version of the Boltzmann code {\tt CLASS} \citep{Lesgourgues:2011re,Blas:2011rf}, based on the birefringence implementation of ref.~\citep{Greco:2022xwj} and\footnote{This implementation exploits the spectetor field approximation, by requiring the axion background energy budget to be negligible with respect to the total energy of the Universe.} the galaxy guidelines of {\tt CLASSgal} \citep{DiDio:2013bqa}.

\subsection{Power spectrum estimation}
\label{subsec:pseudo}
We derive the observed cross-correlation angular power spectrum from the data of section~\ref{sec:data} through a \emph{pseudo}-$C_\ell$ estimator \citep{Hivon:2001jp} at $\ell>12$, whilst exploiting a quadratic maximum likelihood (QML) approach for the first multipoles.\footnote{During the process, we take into account deprojection effects by applying a systematic-template for Quaia quasars to the estimates, following ref.~\citep{Storey-Fisher:2023gca}. We find negligible variations, as also deeply studied in \citep{Alonso:2024emk}. Furthermore, we minimize dipole leakage on the derived spectra by adding a constant template to the covariance.}

\paragraph{\emph{Pseudo}-$C_\ell$.} This estimator works in harmonic space and, taking advantage of its extremely fast computation times, is particularly suited for large datasets. Nonetheless, dealing with spherical harmonic transforms of masked data introduces information loss and mode mixing, especially at the largest scales. Following the {\tt NaMaster} \citep{Alonso:2018jzx} implementation the pseudo angular power spectrum is
\be
    \tilde{C}^{\alpha g}_{\ell}  = \frac{1}{2\ell+1}\sum_{m=-\ell}^{+\ell} \tilde{a}^\alpha_{\ell m}\;\tilde{a}^{g*}_{\ell m} \,,
\ee
where the harmonic coefficients for a generic field $X({\bf\hat{n}})$ come from its expansion on spherical harmonics over a masked region of the sky, defined by the weight function $w^X_{\bf\hat{n}}$,
\be
    \tilde{a}^X_{\ell m} = \int\de\Omega\; X({\bf\hat{n}})\, w^X_{\bf\hat{n}}\, Y^*_{\ell m}({\bf\hat{n}}) \,.
\ee
The unbiased estimator of the true power spectrum for our cross-correlation\footnote{Let us notice how eq.~\eqref{eq:pseudo} does not include a noise power spectrum, as the latter only appears for auto-correlations.} is then:
\be\label{eq:pseudo}
    \hat{C}^{\alpha g}_\ell=\sum_{\ell'}M^{-1}_{\ell \ell'}\tilde{C}^{\alpha g}_{\ell'} \,,
\ee
with $M^{-1}_{\ell \ell'}$ accounting for the incomplete sky coverage
\be
\begin{split}\label{eq:mpseudo}
    M^{\alpha g}_{\ell \ell'} & =  \frac{(2 \ell' + 1)}{ 4 \pi}\sum_{\ell'' } (2 \ell'' + 1)\;\tilde W^{\alpha g}_{\ell''}\;
{\left ( \begin{array}{ccc}
        \ell & \ell' & \ell''  \\
        0  & 0 & 0
       \end{array} \right )^2} \,,\\
    \tilde W^{\alpha g}_\ell &= \frac{1}{2 \ell + 1} \sum_{m=-\ell}^{\ell} \tilde w^{\alpha}_{\ell m}\; \tilde w^{g*}_{\ell m} \,.
\end{split}
\ee
Here, the last term in the first line is a Wigner $3\,$-$\,j$ symbol \citep{Varshalovich:1988ifq}, whilst $w^{\alpha}_{\ell m}$ and $w^{g}_{\ell m}$ are the spherical harmonic coefficients of the two probes' masks, respectively. Due to the relatively strong bias, at large scales, brought in by eq.~\eqref{eq:mpseudo}, we trust only estimates for $\ell>12$, up to $\ell_{\rm max} = 3N_{\rm side}-1$.

\paragraph{QML.} This estimator \citep{Tegmark:1996qt,Tegmark:2001zv}, on the contrary, works in pixel space, sensibly reducing masking effects at low multipoles. However, dealing with several pixel-pixel matrices is computationally demanding. The underlying algebra yields
\be\label{eq:qml}
    \hat C_{\ell}^{\alpha g} = F^{-1}_{\ell \ell'} \, y_{\ell'} \,,
\ee
where the Fisher matrix $F_{\ell\ell'}$ and the minimum-variance quadratic estimator $y_\ell$ are defined as follows:
\be
\begin{split}
    F_{\ell \ell^{\prime}} &= {\rm Tr}\left[\bs{C}^{-1}_{\alpha\alpha} \,\bs{P}_{\ell} {\bf C}^{-1}_{gg} \,\bs{P}_{\ell'}^{T}\right] \,,\\
    y_{\ell} &= \bm{\alpha}^T \bs{C}^{-1}_{\alpha\alpha} \, \bs{P}_{\ell} \bs{C}^{-1}_{gg} \bm{\delta_g} \,.
\end{split}
\ee
$\bs{C}_{\alpha\alpha}$ and $\bs{C}_{gg}$ are the signal covariance matrices, related to the two data vectors in pixel space $\bm{\alpha}$ and $\bm{\delta_g}$. The matrix elements of $\bs{P}_{\ell}$ are ${\rm P}_{\ell,\,ij}=(2\ell+1)/4\pi\;p_\ell({\bf\hat{n}}_i\cdot{\bf\hat{n}}_j)$, defined through the Legendre polynomials $p_\ell({\bf\hat{n}}_i\cdot{\bf\hat{n}}_j)$. Eq.~\eqref{eq:qml} is an unbiased estimator of the underlying cross-correlation ($\langle \hat{C}_\ell^{\alpha g}\rangle=C_\ell^{\alpha g}$), and we utilize it to derive the power spectrum at $\ell\le12$.

\subsection{Covariance estimation}
\label{subsec:cov}
We estimate the covariance matrix of the cross-correlation as
\be\label{eq:cov}
    \mathsf{\hat C}^{\alpha g}_{\ell\ell'} = \frac{1}{n-1}\sum_{k=1}^n \left(\hat C_\ell^{\alpha g(k)} - \langle \hat C_\ell^{\alpha g(k)}\rangle\right) \left(\hat C_{\ell'}^{\alpha g(k)} - \langle \hat C_{\ell'}^{\alpha g(k)}\rangle\right) \,,
\ee
with $n=400$ realizations of anisotropic birefringence and galaxy overdensity outlined in section~\ref{sec:data}. It can be demonstrated that this provides an unbiased estimate of the true covariance matrix $\mathsf{\Sigma}$ as long as $p < n-1$, with $p$ being the length of the data vector $\hat{C}_\ell^{\alpha g}$. However, for our analysis, we require the inverse covariance matrix for both goodness-of-fit assessments and parameter estimation (see section~\ref{sec:res}). As shown in ref.~\citep{Hartlap:2006kj}, $(\mathsf{\hat{C}}^{\alpha g}_{\ell\ell'})^{-1}$ is a biased estimator of the true inverse covariance ${\mathsf{\Sigma}}^{-1}$ due to noise injected in $\mathsf{\hat{C}}^{\alpha g}_{\ell\ell'}$ as $p$ approaches $n$. To first order, this issue can be mitigated by either re-normalizing the estimated inverse covariance \citep{Hartlap:2006kj} or marginalizing over the true covariance matrix, conditioned on its estimate \citep{Sellentin:2015waz}. Furthermore, binning the data vector helps to reduce the noise in the off-diagonal elements of the estimated covariance by compensating for the limited number of realizations. 
Ideally, the data vector length should be reduced to $\tilde{p}$, the maximum size that still allows for reliable covariance estimation given the number of simulations $n$, satisfying $\tilde{p}(\tilde{p} + 1)/2 < n$. For 400 realizations, this yields $\tilde{p} \sim 27$. In what follows, we will adopt the least aggressive binning scheme possible and further examine the impact of binning in appendix~\ref{app:A}.

\begin{figure}[ht]
\centering
\includegraphics[width=0.99\hsize]{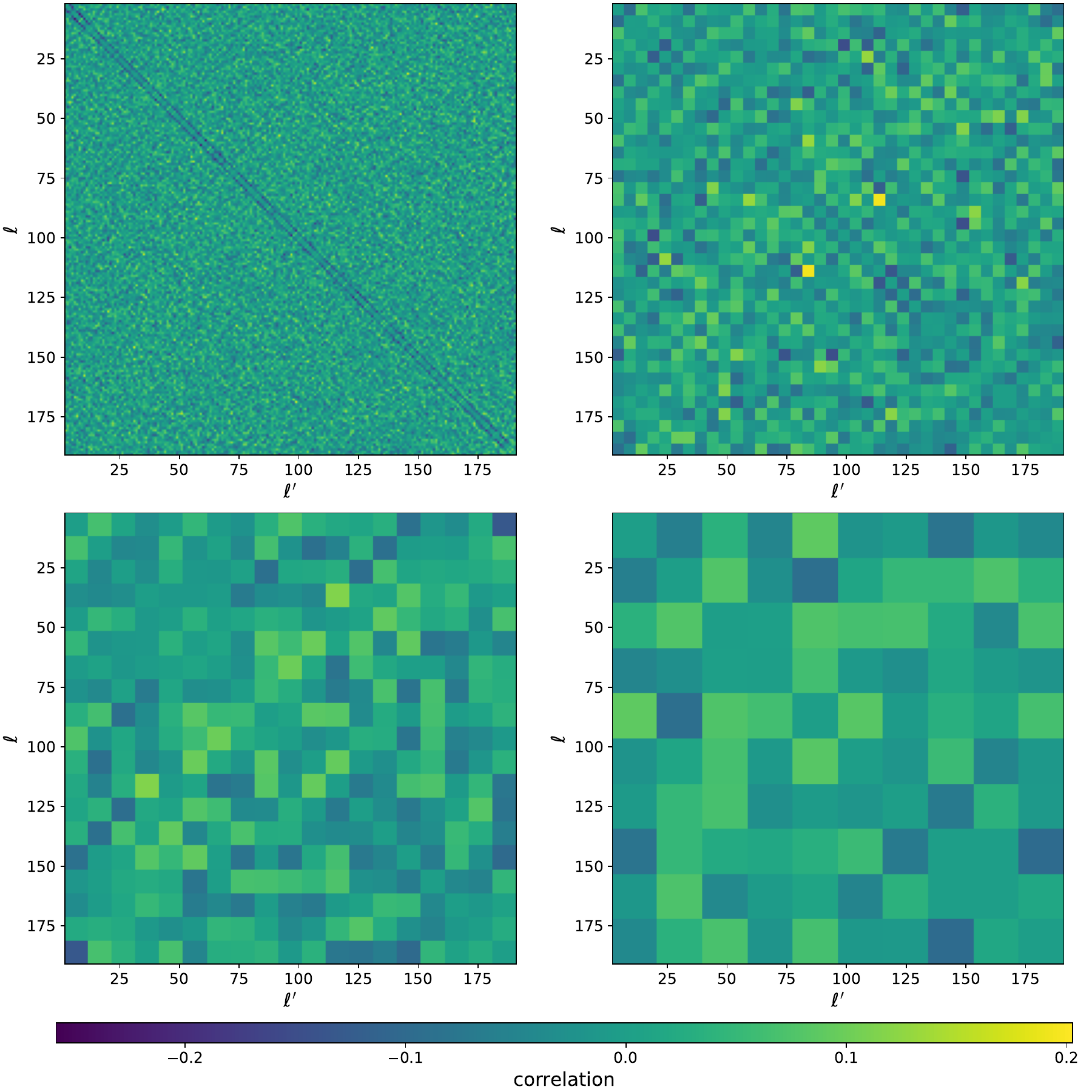}
\caption{Correlation matrix obtained from the covariance of eq.~\eqref{eq:covcorr} with the diagonal subtracted. We exploit a QML estimator for the first 12 multipoles, and \emph{pseudo}-$C_\ell$'s afterwards (see section~\ref{subsec:pseudo}). (\emph{Top left}): no binning, $\Delta\ell=1$. (\emph{Top right}): 5 multipoles per bandpower, $\Delta\ell=5$ (38 bins). (\emph{Bottom left}): 10 multipoles per bandpower, $\Delta\ell=10$ (19 bins). (\emph{Bottom right}): 19 multipoles per bandpower, $\Delta\ell=19$ (10 bins).}
\label{fig:cov}
\end{figure}

Additionally, the random catalog utilized to extract the 400 realizations of galaxy overdensity lacks the large-scale systematics present in the associated data. To address this, we calibrate the effective degrees of freedom, $\nu_{\rm eff}$, from the aforementioned realizations and subsequently re-normalize the covariance of eq.~\eqref{eq:cov} by the variance analytically derived from the data with the calibrated $\nu_{\rm eff}$. This process can be summarized as follows
\be\label{eq:covcorr}
\begin{split}
    \mathsf{\hat C}^{\alpha g}_{\rm new} &= \mathsf{\hat C}^{\alpha g}\frac{\sigma_{\rm dat,\,eff}\otimes\sigma_{\rm dat,\,eff}}{\sigma_{\rm sim}\otimes\sigma_{\rm sim}} \,,\\
    \sigma_{\rm sim} &= \sqrt{{\rm diag}(\mathsf{\hat C}^{\alpha g})} \,,\\
    \sigma_{\rm dat,\,eff} &= \sqrt{\frac{1}{\nu_{\rm eff}}\left((\hat C_{\ell,\,\rm dat}^{\alpha g})^2+\hat C_{\ell,\,\rm dat}^{\alpha \alpha}\hat C_{\ell,\,\rm dat}^{gg}\right)} \,,\\
    \nu_{\rm eff} &= \frac{(\hat C_{\ell,\,\rm sim}^{\alpha g})^2+\hat C_{\ell,\,\rm sim}^{\alpha \alpha}\hat C_{\ell,\,\rm sim}^{gg}}{\sigma_{\rm sim}^2} \,,
\end{split}
\ee
where $\sigma_{\rm sim}$ represents the standard deviation derived from eq.~\eqref{eq:cov} exploiting the 400 realizations of the underlying spectra, with their means denoted as $\hat C_{\ell,\,\rm sim}^{\alpha g}$, $\hat C_{\ell,\,\rm sim}^{\alpha \alpha}$ and $\hat C_{\ell,\,\rm sim}^{gg}$. Meanwhile, $\sigma_{\rm dat,\,eff}$ refers to the analytical standard deviation calculated from the data ($\hat C_{\ell,\,\rm dat}^{\alpha g}$, $\hat C_{\ell,\,\rm dat}^{\alpha \alpha}$ and $\hat C_{\ell,\,\rm dat}^{gg}$), adjusted using the effective degrees of freedom $\nu_{\rm eff}$. The $\otimes$ product refers to the outer product to compute the correlation matrix $\rho$ related to both simulations and data ($\rho = \sigma\otimes\sigma$). The covariance structure is illustrated in Fig.~\ref{fig:cov}, where we present $\mathsf{\hat C}^{\alpha g}_{\rm new}/\sqrt{{\rm diag}(\mathsf{\hat C}^{\alpha g}_{\rm new})}\otimes\sqrt{{\rm diag}(\mathsf{\hat C}^{\alpha g}_{\rm new})}$, excluding the diagonal elements, for four different binning schemes ($\Delta\ell=1$, $\Delta\ell=5$, $\Delta\ell=10$, $\Delta\ell=19$). This representation is valuable for analyzing the off-diagonal contributions and shows how binning reduces correlations coming from the off-diagonal structure. For the least aggressive cases, the initial off-diagonal terms exhibit mild anti-correlation.

\section{Results}
\label{sec:res}
The focus of this paper is to present, for the first time, a measurement of the cross-correlation between anisotropic birefringence and galaxies. Both probes stem from the metric perturbations and a detection of such a signal would not only rise valuable insights about cosmic birefringence, but also allow to constrain the underlying axion-parameters \citep{Arcari:2024nhw}. Exploiting a combined QML/\emph{pseudo}-$C_\ell$ analysis we derive the observed angular power spectrum from {\it Planck} NPIPE polarization data and Quaia galaxy overdensity. We show how this signal is well compatible with the null-hypothesis and compare it with the theoretical power spectrum to impose a novel bound on the axion-photon coupling for ultralight axions.

\subsection{The measured cross-correlation}
\label{subsec:cross}
\begin{figure}[t]
\centering
\includegraphics[width=0.95\hsize]{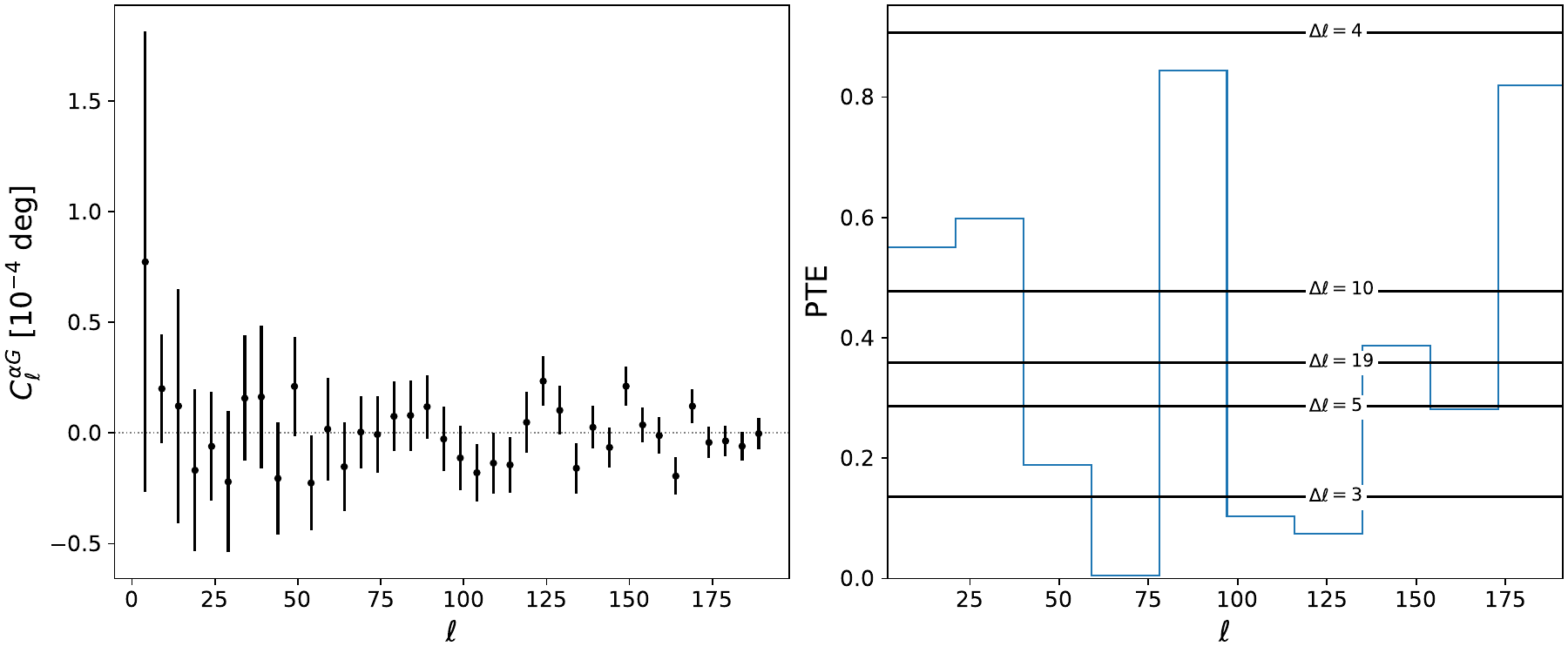}
\caption{(\emph{Left}): cross-correlation angular power spectrum between anisotropic birefringence and galaxy number counts, as measured by the combination of {\it Planck} NPIPE polarization data and the quasar catalog Quaia. The signal is binned with $\Delta\ell=5$ bandwidth and error bars are computed from the covariance estimated with 400 realizations of the two probes (see section~\ref{subsec:cov}). (\emph{Right}): the black horizontal lines corresponds to the PTE obtained across the entire range of multipoles with different binning schemes as indicated by the related bandwidth $\Delta\ell$. The blue line represents the PTE achievable within a specific chunk of multipoles and $\Delta\ell=1$.}
\label{fig:spectra}
\end{figure}
As discussed in section~\ref{sec:meth}, we estimate the underlying cross-correlation angular power spectrum utilizing a QML estimator for $\ell=[2,12]$ and a \emph{pseudo}-$C_\ell$ estimator for $\ell=[13,191]$.\footnote{Let us remind that this choice is due the inefficiency of \emph{pseudo}-$C_\ell$'s to correctly estimate power spectra at large scales, where masking effects in harmonic space are dominant.} The maximum multipole of our analysis stems from the resolution of the datasets under consideration ($\ell_{\rm max}=3N_{\rm side}-1$ with $N_{\rm side}=64$). The estimators are applied to the maps, and related masks, presented in section~\ref{sec:data}. Anisotropic birefringence is associated to the {\it Planck} NPIPE polarization dataset, whilst the spatial distribution of galaxies is derived from the Quaia quasar catalog. 

The result is shown in the left panel of fig.~\ref{fig:spectra} with 5-multipoles per bandpower. The associated error bars spring from the diagonal of the covariance in eq.~\eqref{eq:covcorr}. This signal is compatible with the null-hypothesis with a probability to exceed (PTE) of 29\%. The compatibility is retained across different binning strategies and shown in the right panel of fig.~\ref{fig:spectra} for $\Delta\ell=3,\,4,\,5,\,10\;{\rm and}\;19$. Only the unbinned and $\Delta\ell=2$ spectrum yield poor compatibility with the null-hypothesis (PTE $\ll 5\%$) due to the high level of correlation in the off-diagonal structure of the covariance (see fig~\ref{fig:cov}). Let us highlight that the following results show little to no change when varying the binning scheme (for further detail on the impact of binning refer to appendix~\ref{app:A}). The right panel of fig.~\ref{fig:spectra} also illustrates how the measured cross-correlation stems compatibility with the null-hypothesis in selected chunks of multipoles.\footnote{Only the chunk around $\ell=60$ results in a low PTE due to the off-diagonal terms of the covariance being the strongest in this region (see fig.~\ref{fig:cov}).} Each of the ten chunks contains 19 multipoles, with a $\Delta\ell=1$ bandwidth. 

\begin{figure}[t]
\centering
\includegraphics[width=0.95\hsize]{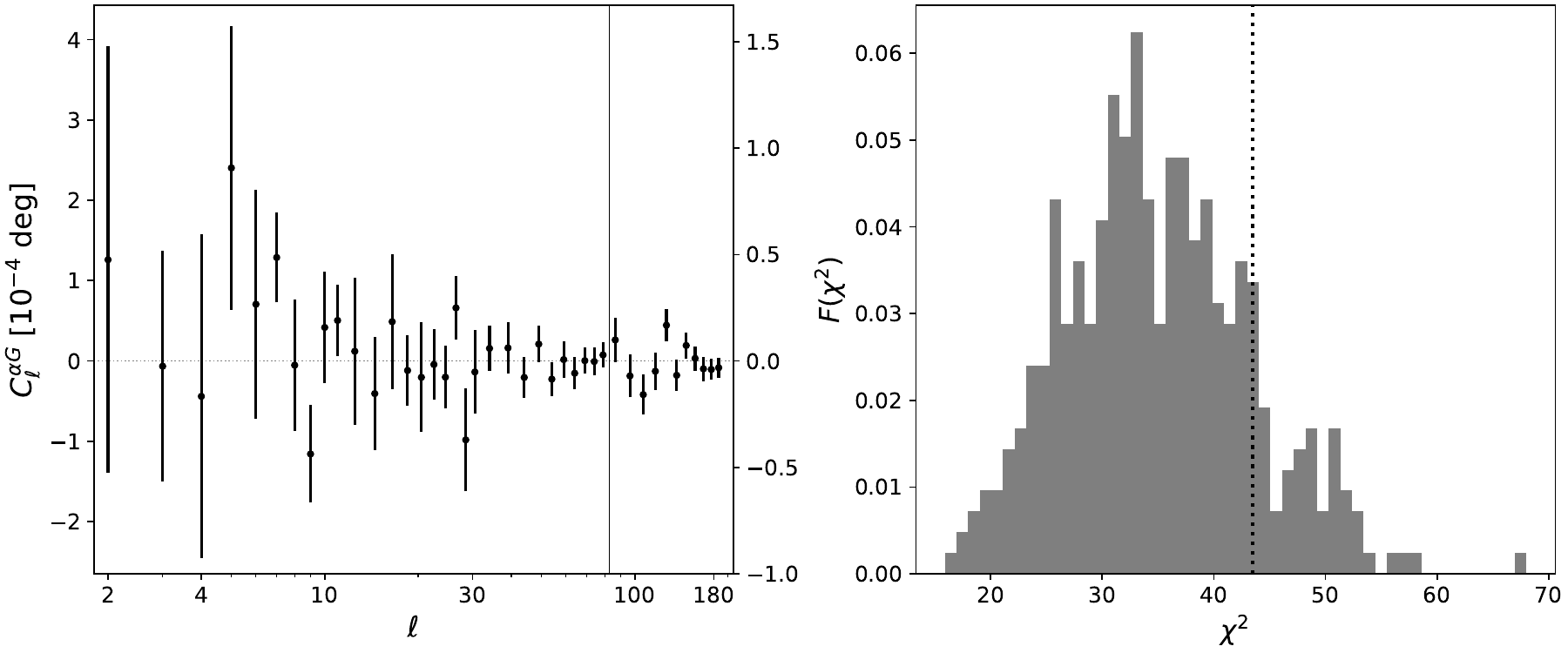}
\caption{(\emph{Left}): Planck NPIPE cross Quaia angular power spectrum between anisotropic birefringence and galaxy number counts. We use the following binning scheme: no binning for the first 10 multipoles, 10 bins with 2 multipoles, 10 bin with 5 multipoles and 10 multiples per bandpower afterwards. The result is divided in two y-scale regions for visualization purposes. (\emph{Right}): Corresponding Gaussian chi-squared distribution from the 400 realizations. The dotted vertical line refers to the data $\chi^2$ of the left panel.}
\label{fig:chi}
\end{figure}

For parameter estimation, in section~\ref{subsec:bound}, we will utilize the following binning scheme: $\Delta\ell = 1$ for the first 10 multipoles, $\Delta\ell = 2$ for the following 20 multipoles, $\Delta\ell = 5$ for the consequent 50 multipoles and $\Delta\ell = 10$ for the rest. This approach minimizes information loss on large scales, where our theoretical analysis predicts the greatest power \citep{Arcari:2024nhw}. This binning strategy remains compatible with zero, with a PTE of 37\%. The left panel of fig.~\ref{fig:chi} shows its angular power spectrum, while the right panel illustrates the corresponding Gaussian null-hypothesis $\chi^2$ for data, overlapped with the distribution given by the 400 realizations used to estimate the covariance matrix.

We emphasize how this is the first measurement of such a correlation and future CMB polarization experiments \citep{SimonsObservatory:2018koc,LiteBIRD:2022cnt,CMB-S4:2020lpa}, together with forthcoming galaxy surveys \citep{Euclid:2019clj,Euclid:2021icp,DESI:2013agm,DESI:2016fyo}, could enhance the signal-to-noise ratio and eventually reveal an excess of power (see e.g. \citep{Arcari:2024nhw}).

\subsection{Scale invariant amplitude}
\label{subsec:amp}
\begin{figure}[t]
\centering
\includegraphics[width=0.95\hsize]{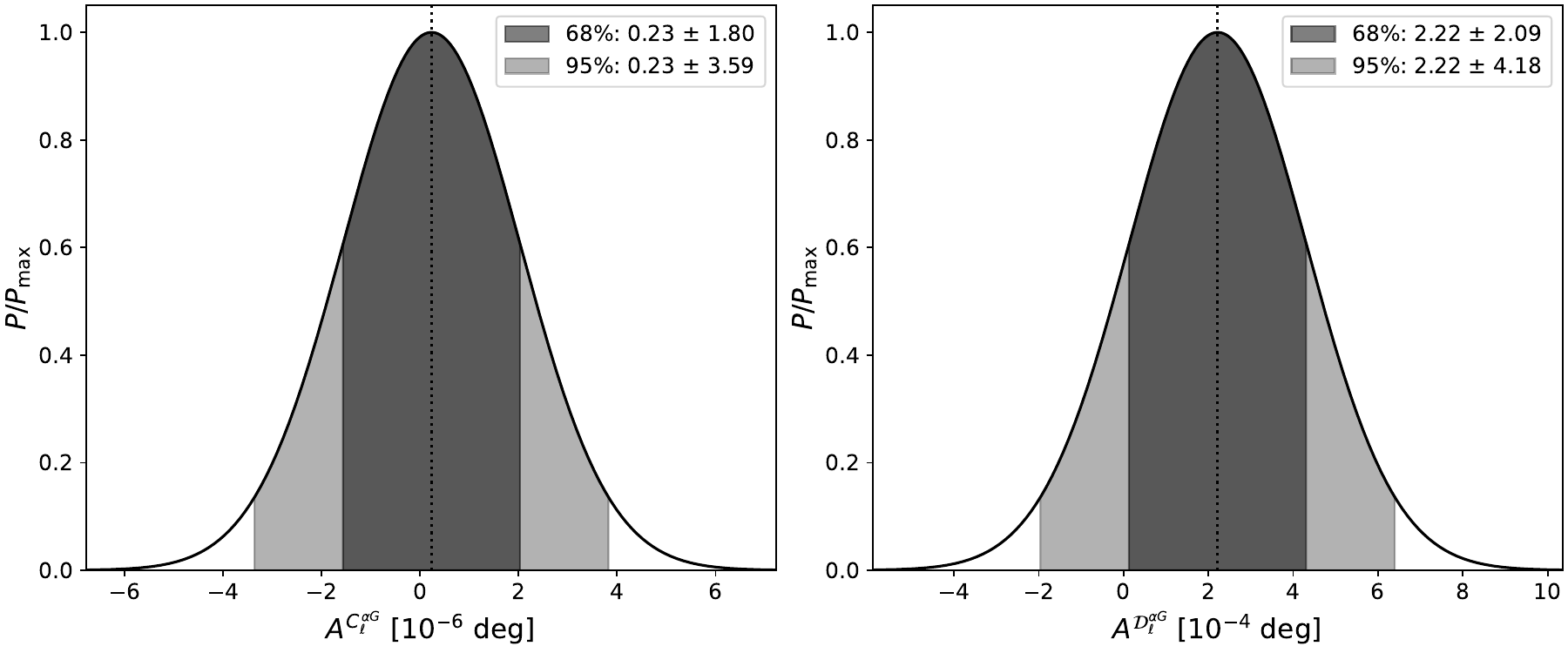}
\caption{Posterior on a scale invariant amplitude $A$, fitted on the observed spectrum of fig.~\ref{fig:spectra} with a Gaussian likelihood, both for $C_\ell$'s (\emph{Left}) and $\mathcal{D}_\ell$'s (\emph{Right}). The dark(light) shaded region corresponds to the 68(95)\% confidence level.}
\label{fig:amp}
\end{figure}
As an extra test of data compatibility with the null-hypothesis, we perform a Gaussian likelihood on the scale-invariant amplitude, $A$, of the $\alpha$-G cross-correlation field:
\be\label{eq:amp}
    -2\,{\rm log}\mathcal{L}(A) = (C_\ell-A)\,\mathsf{(C^{-1})}_{\ell\ell'}\,(C_{\ell'}-A)^T \,,
\ee
where the superscript $\alpha g$ is implied and $\mathsf{(C^{-1})}_{\ell\ell'}$ is the inverse covariance matrix derived from eq.~\eqref{eq:covcorr}. From eq.~\eqref{eq:amp} we derive the best fit value of the amplitude and the related 68\% and (95\%) confidence interval (we use a 5-multipole per bandpower binning scheme): $A^{C_\ell}=[0.23\pm1.80(3.59)]\times10^{-6}\,\deg$. We also fit a constant amplitude on $\mathcal{D}_\ell=\ell(\ell+1)/2\pi\,C_\ell$, by replacing it to $C_\ell$ both for the data vector and the covariance of eq.~\eqref{eq:amp}. We obtain $A^{\mathcal{D}_\ell}=[2.22\pm2.09(4.18)]\times10^{-4}\,\deg$. The two results are depicted in fig.~\ref{fig:amp}. Additionally, we calculated both values using an unbiased inverse covariance following Hartlap et al.~\citep{Hartlap:2006kj}, and by adjusting the likelihood in eq.~\eqref{eq:amp} according to the method in ref.~\citep{Sellentin:2015waz}, which marginalizes over the true covariance matrix. As discussed in section~\ref{subsec:cov}, these approaches account for the limited number of realizations and broaden the posterior distribution. Consequently, the resulting confidence levels are increased by approximately 3\% (the effect would be larger for less aggressive binning schemes).

\subsection{Bound on axion-photon coupling}
\label{subsec:bound}
\begin{figure}[t]
\centering
\includegraphics[width=0.55\hsize]{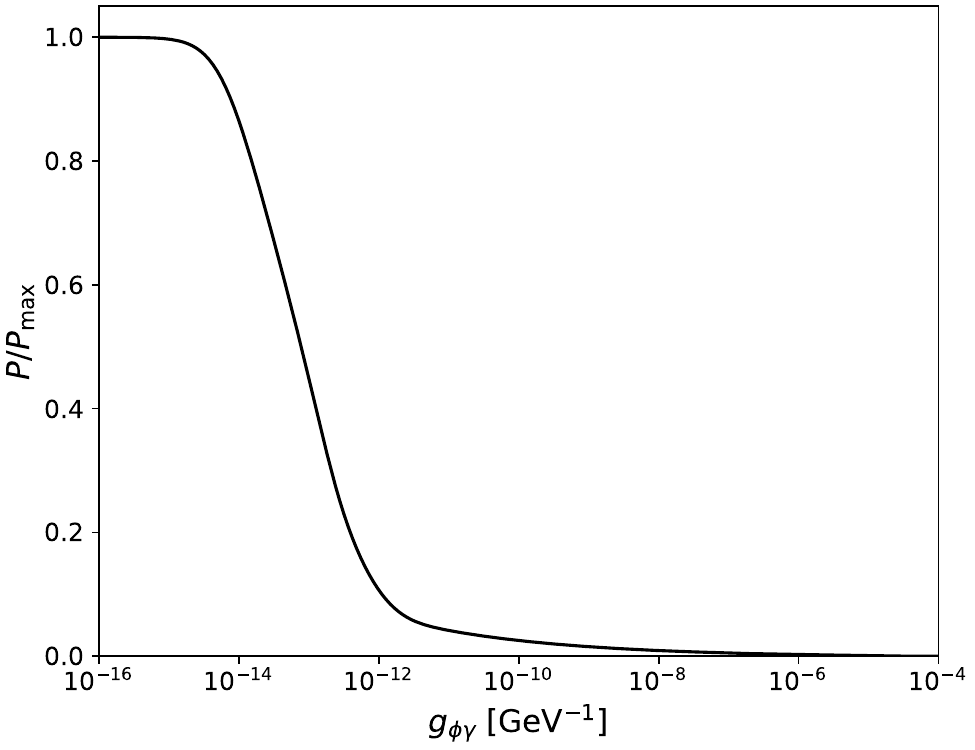}
\caption{Posterior distribution on the axion-photon coupling $g_{\phi\gamma}$ marginalized on the other two free parameters (mass and initial misalignment). As the coupling increases, the predicted spectrum from eq.~\eqref{eq:alphaG} becomes larger, eventually deviating from observations and thus resulting in a lower probability.}
\label{fig:post}
\end{figure}
In section~\ref{subsec:theory} we have reviewed the theoretical formulation of the birefringence-galaxy cross-correlation and its implementation in our modified version of {\tt CLASS}. In particular, we have discussed how the latter depends on the axion mass $m_\phi$, the initial misalignment $\theta_i$ and the axion-photon coupling $g_{\phi\gamma}$. As extensively outlined in refs.~\citep{Arcari:2024nhw,Greco:2022xwj,Greco:2024oie}, ultralight axions with $m_\phi<10^{-28}\,\si{\electronvolt}$ offer as a promising tool to constrain birefringence and can provide a suitable explanation~\cite{Greco:2024oie} for the latest measurements of the isotropic CB angle \citep{Eskilt:2022wav,Diego-Palazuelos:2022cnh}. While the mass and the misalignment mainly govern the shape of the predicted cross-correlation signal, due to their non trivial behavior within eq.~\eqref{eq:alphaG}, the axion-photon coupling controls directly the amplitude, appearing as a multiplicative factor. We anticipate a probability close to unity for all "small enough" couplings, given the measured power spectrum's compatibility with the null-hypothesis, whilst as the coupling increases the theoretical model will exceed observations and eventually an upper bound on $g_{\phi\gamma}$ can be imposed. Previous works have constrained this coupling through haloscopes, helioscopes, colliders, and astrophysical searches (for comprehensive data and references refer to {\tt AxionLimits} \citep{axionlimits}). These constraints extend down to $m_\phi \sim 10^{-12}\,\si{\electronvolt}$, with extensions to $10^{-24}\,\si{\electronvolt}$ from isotropic birefringence \citep{Galaverni:2009zz,Galaverni:2023zhv} and black hole polarimetry \citep{Gan:2023swl}. Ref.~\citep{Greco:2024oie} probes a combination of axion parameters in the mass range $m_\phi\in\left(10^{-33},\;10^{-26}\right)\,\si{\electronvolt}$, making use of the isotropic birefringence angle from ref.~\citep{Diego-Palazuelos:2022dsq}. Our analysis targets an ultra-low mass regime where, to date, no direct constraint on the axion–photon coupling has been reported in the literature.

We explore the axion-parameter space with a Gaussian likelihood
\be\label{eq:like}
    -2\,{\rm log}\mathcal{L}(\Theta) = (C_\ell^{\rm th}(\Theta)-C_\ell^{\rm obs})\,\mathsf{(C^{-1})}_{\ell\ell'}\,(C_\ell^{\rm th}(\Theta)-C_\ell^{\rm obs})^T \,,
\ee
where $C_\ell^{\rm th}$ is the predicted power spectrum from the theoretical model of eq.~\eqref{eq:alphaG} with parameters $\Theta=(m_\phi,\,\theta_i,\,g_{\phi\gamma})$, $C_\ell^{\rm obs}$ the observed power spectrum as discussed in section~\ref{subsec:cross}, and $\mathsf{(C^{-1})}_{\ell\ell'}$ the inverse of the covariance matrix of eq.~\eqref{eq:covcorr}. We consider $m_\phi\in\left[10^{-33},10^{-28}\right]~\si{\electronvolt}$,\footnote{The lowest mass roughly corresponds to the Hubble parameter today and even lower masses would lead to fields whose dynamical evolution has not commenced yet. We do not explore the region where $m_\phi>10^{-28}\,\si{\electronvolt}$ as it is computationally demanding and the perturbative approach breaks \citep{Arcari:2024nhw}.} and $\theta_i\in\left(-\pi,\,\pi\right)$, whereas we let the coupling vary, in principle, between 0 and $\infty$. 
\begin{figure}[t]
\centering
\includegraphics[width=0.65\hsize]{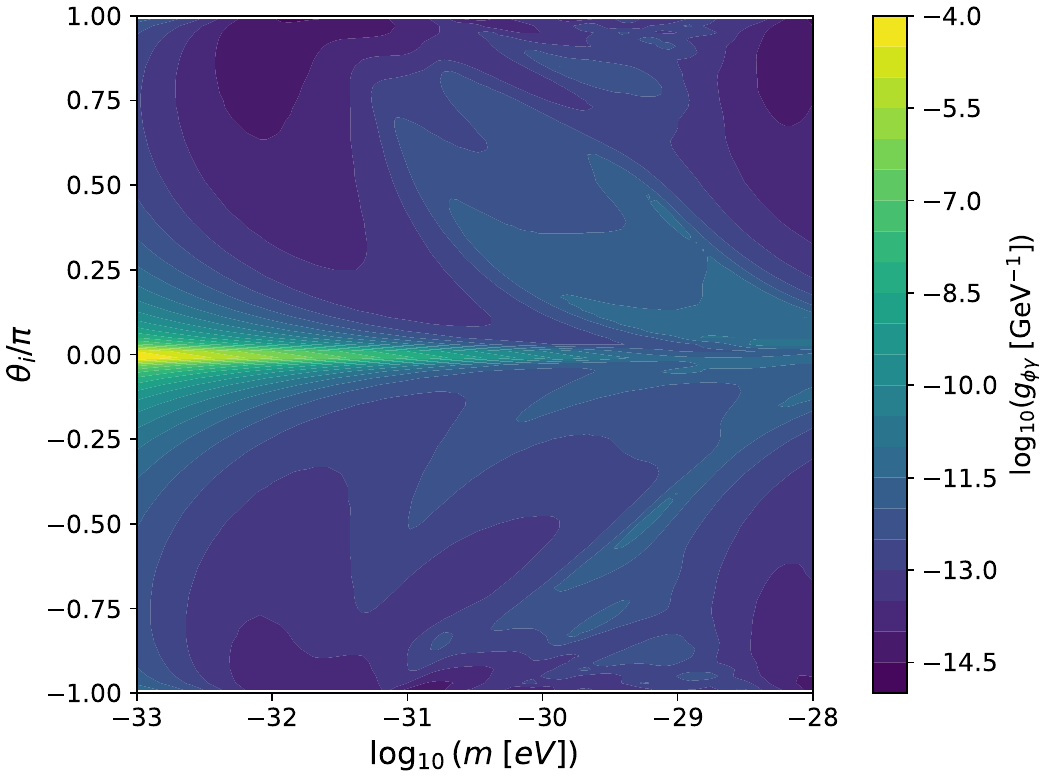}
\caption{Achievable upper bound on the axion-photon coupling, at the 95\% confidence level, for each point in the $m_\phi$-$\theta_i$ parameter space. For very small misalignment angles, the theoretical power spectrum is significantly reduced, requiring a higher value of $g_{\phi\gamma}$ to enhance the signal sufficiently away of observations.}
\label{fig:bound}
\end{figure}

\begin{figure}[t]
\centering
\includegraphics[width=0.95\hsize]{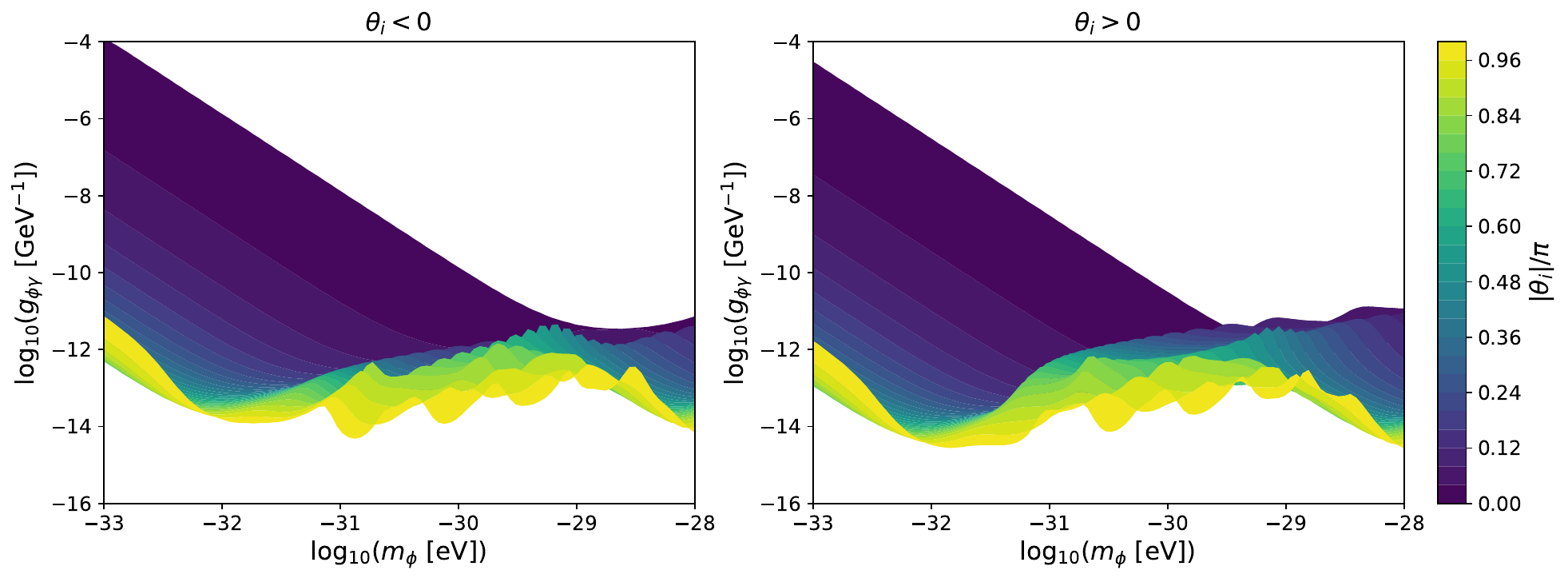}
\caption{Upper bound, at $95 \%$ confidence level, on the axion-photon coupling as a function of the axion mass. Each line is drawn for a fixed value of the initial misalignment angle, depicted in the colorbar. It is evident how the tightest constraints are obtained for high values of the latter, where there is no degeneracy with the coupling.}
\label{fig:bound_m}
\end{figure}

\begin{figure}[t]
\centering
\includegraphics[width=0.55\hsize]{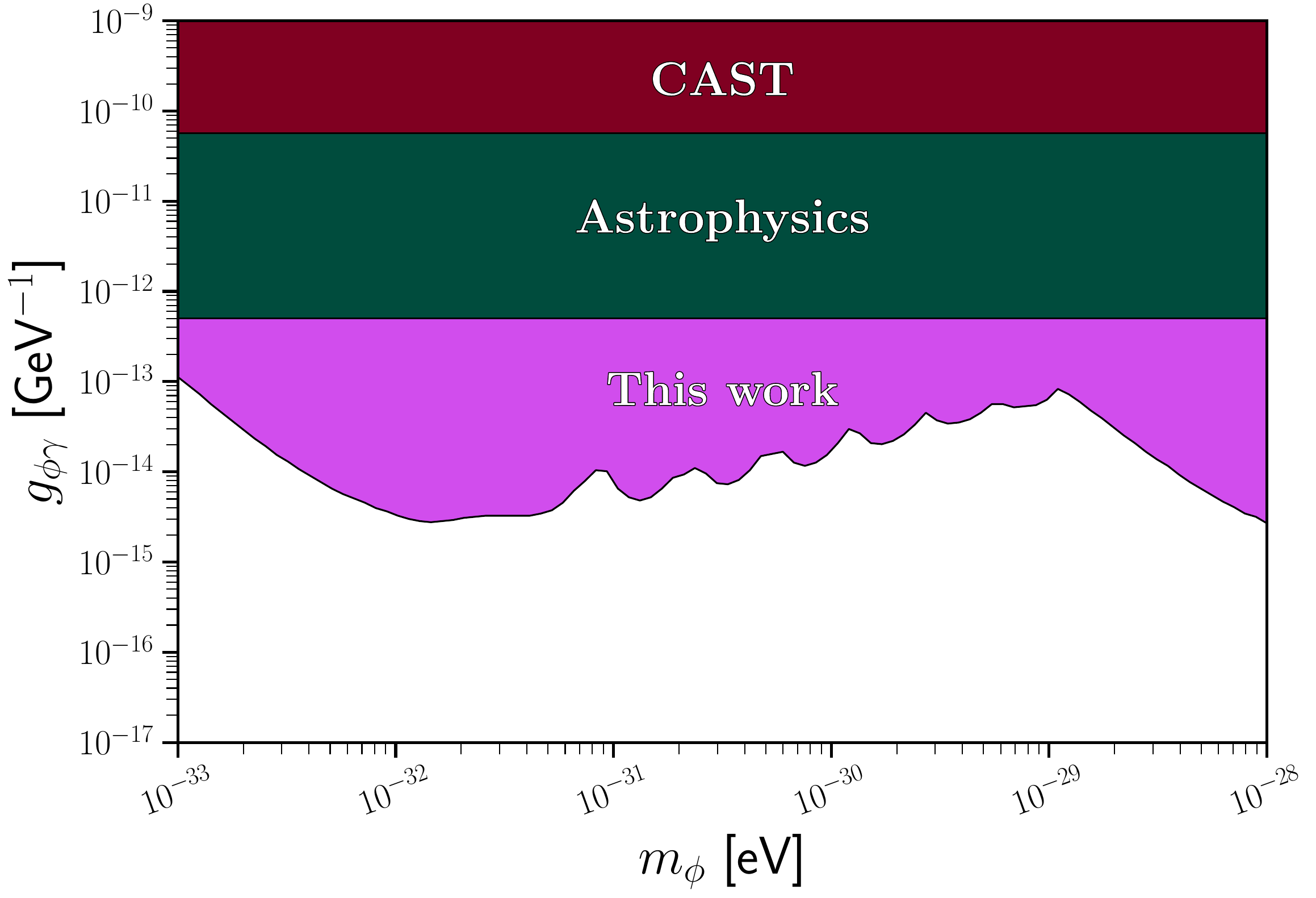}
\caption{Best achievable upper bound, at $95 \%$ confidence level, on the axion-photon coupling, $g_{\phi\gamma}$, as a function of the axion mass, $m_\phi$, as derived from the cross-correlation analysis of {\it Planck} PR4 birefringence and Quaia galaxy number counts. We compare the result with the extension to lower masses of current bounds, derived in a higher mass range. \citep{axionlimits}.}
\label{fig:ouraxionlimits}
\end{figure}

The marginalized posterior distribution on $g_{\phi\gamma}$ is shown in fig.~\ref{fig:post}. The region of highest probability contains couplings smaller than $\sim10^{-12}\,\si{\giga\electronvolt^{-1}}$, but a significant portion of the probability density extends to the tail of the distribution. At the 95\% confidence level, the upper bound reaches over $10^{-5}\,\si{\giga\electronvolt^{-1}}$. This is primarily due to a degeneracy with the initial misalignment, which also determines the starting value of the axion field. Consequently, when $|\theta_i|\sim0$, the cross-correlation is null regardless of the coupling value. To address this, we derive an upper bound for each point in the $m_\phi$-$\theta_i$ parameter space, presenting our results in fig.~\ref{fig:bound}. The regions with the strongest upper bounds are those where $m_\phi\sim10^{-32}\,\si{\electronvolt}$ and the misalignment angle is high. This result aligns with our previous theoretical analysis \citep{Arcari:2024nhw}. In fig.~\ref{fig:bound_m}, we also show the corresponding curves in the coupling-mass plane at fixed values of $\theta_i$. It is clear that the closer the initial misalignment is to zero, the less stringent the obtained bound becomes. In conclusion, cross-correlating anisotropic birefringence with galaxies using the currently available data allows us to set an upper bound on the axion-photon coupling down to $\sim10^{-15}\,\si{\giga\electronvolt^{-1}}$ in the most favorable region of our parameter space.
This makes for an unprecedented result in the ultralight axion mass range under consideration, significantly improving upon existing astrophysical and laboratory bounds at higher masses\footnote{These upper bounds are formally valid for $m_\phi>10^{-24}\,\si{\electronvolt}$ and have been linearly extrapolated into our mass range solely for comparison purposes.} (see fig.~\ref{fig:ouraxionlimits}). 

\section{Conclusion}
\label{sec:conc}
In this study, we have presented, for the first time, the measurement of the cross-correlation between anisotropic cosmic birefringence and the spatial distribution of galaxies. The rotation of the linear polarization plane of the CMB can be caused by axion-like components coupling to photons with parity-violating terms. The evolution of these pseudo-scalars is governed by the interplay between their potential and metric perturbations, the latter being responsible for the collapse of structures into forming galaxies. This cross-correlation signal is not only expected to be non-zero but also serves as a valuable probe of the underlying axion model.

We computed the observed power spectrum by combining {\it Planck} NPIPE polarization data with the Quaia quasar catalog. By applying an EB-estimator on the former, we obtained a birefringence map across the sky at $N_{\rm side}=64$. Similarly, we calculated the corresponding galaxy overdensity map using the 1.3 million sources from the latest Quaia release. We applied a QML estimator at $\ell<12$, integrated with a \emph{pseudo}-$C_\ell$ estimator at larger multipoles. We computed the related covariance matrix from 400 realizations of the two probes and showed that our measurement is well compatible with zero, with a PTE between 20\% and 80\% depending on the binning scheme. This compatibility is also retained across selected chunks of the entire multipole range.

With the aim of setting an unprecedented upper bound on the axion-photon coupling for ultralight candidates with masses in the range $(10^{-33},\,10^{-28})\,\si{\electronvolt}$, we performed a Gaussian likelihood analysis on the following free parameters: the mass $m_\phi$, the initial misalignment angle $\theta_i$, and the axion-photon coupling $g_{\phi\gamma}$. Due to a degeneracy of the coupling with the very small end of the initial misalignment, we were unable to retrieve a competitive upper bound with the full normalization of the other two parameters. Nonetheless, we explored the attainable bound at each point in the mass-misalignment parameter space and found values down to $10^{-15}\,\si{\giga\electronvolt^{-1}}$.

In light of the incoming data from next-generation galaxy surveys and CMB polarization experiments, the novelty of this probe lies in its capability to effectively constrain the axion-parameter in a poorly bounded mass range. Additionally, a detection of such a signal could significantly contribute to the open question of the existence of cosmic birefringence and the underlying parity-violating physics.

\acknowledgments
We thank M. Ballardini, L. Caloni and G. Galloni for valuable discussions. We also convey our appreciation towards the Institute for Fundamental Physics of the Universe (IFPU), that hosted the focus week "Parity violation through CMB observations"\footnote{\url{https://www.ifpu.it/focus-week-2024-05-27/}}, where the many insights enabled us to get this work started. We acknowledge financial support from the INFN InDark initiative and from the COSMOS network (www.cosmosnet.it) through the ASI (Italian Space Agency) Grants 2016-24-H.0, 2016-24-H.1-2018 and 2020-9-HH.0.
We acknowledge the use of CINECA HPC resources from the InDark project in the framework of the INFN-CINECA agreement. 
SA and ML acknowledge participation in the COST Action CA21106 "COSMIC WISPers", supported by COST (European Cooperation in Science and Technology). SA and GZ acknowledge financial support by “Bando Giovani anno 2023 per progetti di ricerca finanziati con il contributo 5x1000 anno 2021”.
NB, AG and PN acknowledges support by the MUR PRIN2022 Project “BROWSEPOL: Beyond standaRd mOdel With coSmic microwavE background POLarization”-2022EJNZ53 financed by the European Union - Next Generation EU.
We express our gratitude to Led Zeppelin for inspiring us on the challenging journey towards uncovering dark components, our cosmological “heaven”.

\appendix
\section{Impact of binning and galaxy bias model}
\label{app:A}
As discussed in section~\ref{subsec:cov}, the covariance matrix in eq.~\eqref{eq:cov} may exhibit excessive noise in its off-diagonal elements due to the limited number of realizations used for estimation, potentially leading to bias in the inverse covariance matrix. Increasing the level of binning helps mitigate this bias by reducing strong or anomalous correlations in off-diagonal terms, though at the cost of some information loss, which may warrant alternative strategies. Hartlap et al.~\citep{Hartlap:2006kj} provide a first-order correction for this bias through a constant (Hartlap) factor, dependent on the length of the data vector $p$ and the number of simulations $n$ used to estimate the covariance:
\be\label{eq:icov_har}
    \mathsf{(\hat C^{-1})}^{\alpha g}_{\ell\ell'} = \frac{n-p-2}{n-1}(\mathsf{\hat C}^{\alpha g}_{\ell\ell'})^{-1} \quad\text{for }p<n-2 \,.
\ee
In this context, the inverse matrix of eq.~\eqref{eq:cov} tends to be increasingly overestimated as the length of the data vector approaches the number of available simulations. Notably, when $p \ll n$, the correction factor in eq.~\eqref{eq:icov_har} is close to unity. In cosmology, this often results in an artificially heightened constraining power for parameter estimation. Applying the Hartlap correction factor addresses this by broadening posterior distributions and thereby increasing confidence intervals on parameter estimates. Sellentin and Heavens~\citep{Sellentin:2015waz} suggest that this adjustment introduces heavy tails in the posterior, and propose a more precise method that involves marginalizing over the true covariance matrix, conditioned on its estimate, thus relaxing the Gaussian assumption in the likelihood. This approach effectively results in a t-distribution, where both the mean and variance are treated as random variables. The modified likelihood thus becomes
\be
    -2\ln\mathcal{L} = n\ln{\left(1+\frac{\chi^2}{n-1}\right)} \,,
\ee
where $\chi^2$ is the standard Gaussian likelihood.

\begin{figure}[t]
\centering
\includegraphics[width=0.452\hsize, valign=t]{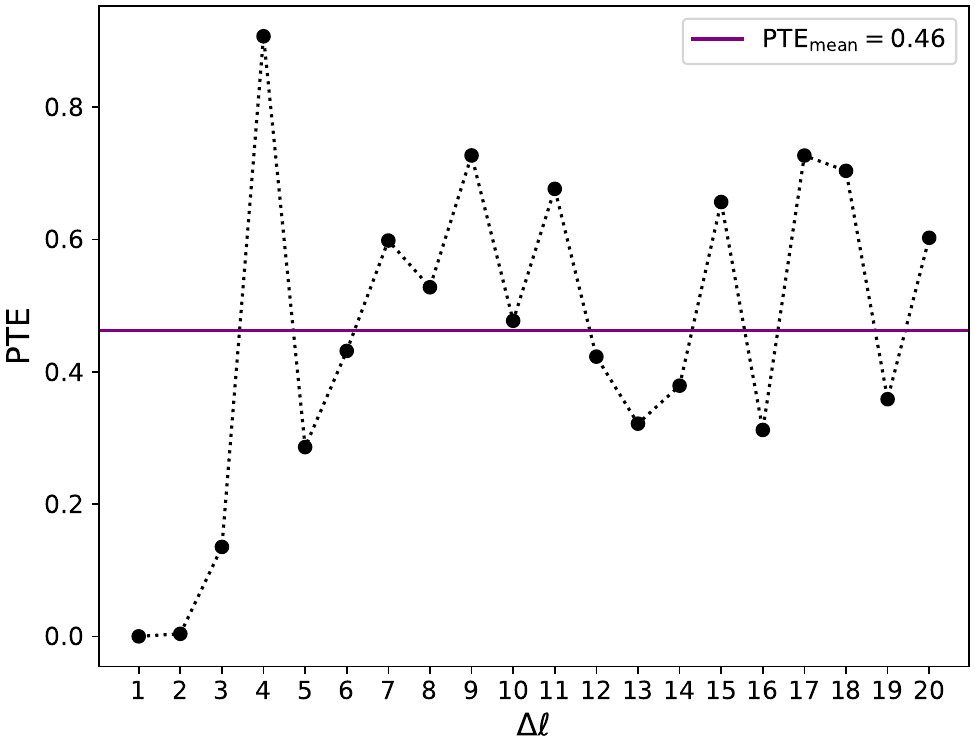}
\includegraphics[width=0.49\hsize, valign=t]{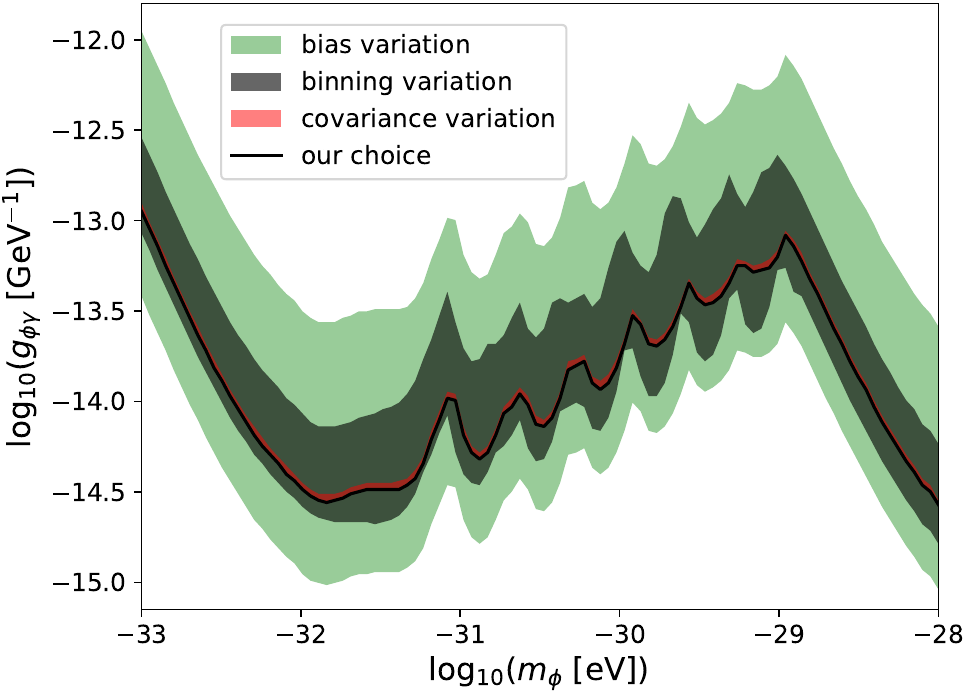}
\caption{(\emph{Left}): value of the PTE as a function of binning bandwidth $\Delta\ell$. Each value corresponds to the compatibility of the measured cross-correlation with zero. (\emph{Right}): maximal variation of the axion-photon coupling, as a function of mass, upon variations of the binning bandwidth (gray) or the covariance estimation and consequent likelihood choice (red). The black line refers to our baseline: no correction applied on the covariance and a mixed binning (10 bins with $\Delta\ell=1$, 10 with $\Delta\ell=2$, 10 with $\Delta\ell=5$ and $\Delta\ell=10$ afterwards).}
\label{fig:ptes}
\end{figure}
\begin{figure}[t]
\centering
\includegraphics[width=0.99\hsize]{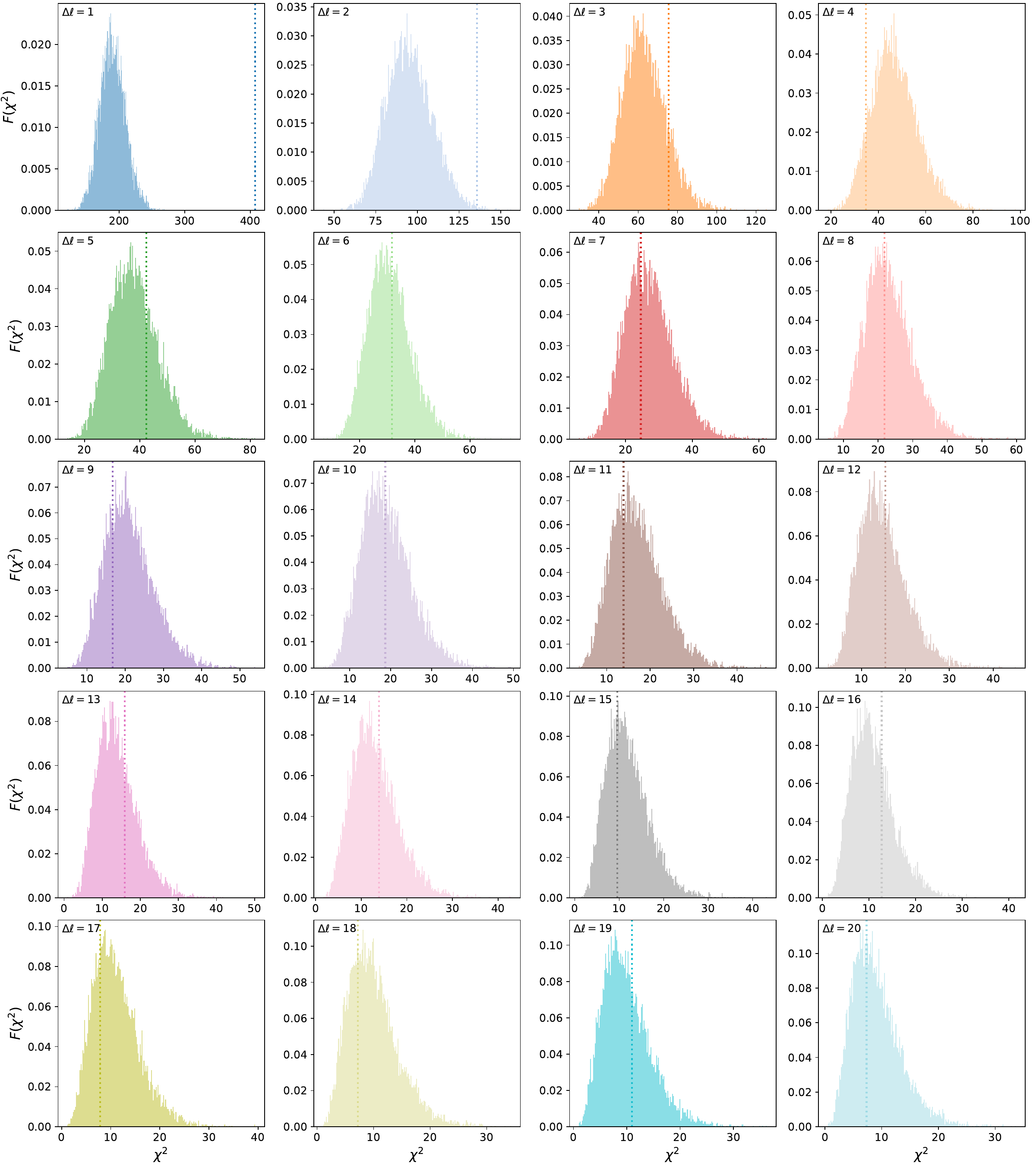}
\caption{Chi-squared distributions for different binning bandwidths $\Delta\ell$, obtained from the 400 realizations. The vertical dotted lines indicates the $\chi^2$ of data.}
\label{fig:chis}
\end{figure}

We evaluated the effects of using a binning scheme and applying each correction on our results. The cross-correlation signal shown in fig.\ref{fig:spectra} remains consistent with zero regardless of the binning choice, and neither the Hartlap factor nor the modified likelihood affects goodness-of-fit tests.\footnote{This occurs because the data’s $\chi^2$ shifts along with the underlying distribution, resulting in an unchanged PTE.} The left panel of fig.~\ref{fig:ptes} shows the PTE of the null-hypothesis as a function of binning bandwidth, $\Delta \ell$. Only the unbinned and 2 multipoles per bin cases produce a value below 5\%, due to significant off-diagonal correlations in the covariance matrix, which are mitigated with increased binning bandwidth. In fig.~\ref{fig:chis}, we also present the $\chi^2$ distribution from 400 realizations, with the data value (marked as a vertical line) well within the distribution for all but the first two cases. 

For parameter estimation, the achievable upper bound on the axion-photon coupling is modestly influenced by the choice of binning scheme. More aggressive binning results in less stringent bounds due to information loss on large scales, where the theoretical signal is expected to be strongest \citep{Arcari:2024nhw}. The maximum observed variation in the best $g_{\phi\gamma}$-$m_\phi$ bound of fig.~\ref{fig:bound_m} reaches about half an order of magnitude. This variation diminishes to a few percent as the binning bandwidth exceeds $\Delta\ell \sim 5$, likely indicating that no further large-scale information is being lost. This outcome is shown in the right panel of fig.~\ref{fig:ptes} as a gray-shaded band, overlaying the black line that represents our selected binning scheme as discussed in section~\ref{sec:res}. The red-shaded region shows the maximal variation when applying either the Hartlap correction or the Sellentin and Heavens modified likelihood, with the former reducing constraining power by approximately 10\%, and the latter by about 1\%.

Additionally, our results are influenced by the choice of galaxy bias, modeled as in eq.~\eqref{eq:bias} and derived from the best-fit parameters of ref.~\citep{eBOSS:2017ozs}. To evaluate the sensitivity of our upper bound to variations in galaxy bias, we introduce a scaling factor $b_g$ and re-define the bias as 
\begin{equation}
    b(z) = b_g\left[0.278\left[(1 + z)^2 - 6.565\right] + 2.393\right] \,.
\end{equation}
Following \citep{Piccirilli:2024xgo}, we adopt a uniform prior $b_g \in [0.1, 3.0]$. This scaling factor, being a constant multiplier of the final cross-correlation spectrum, is partially degenerate with the axion-photon coupling. Marginalizing over the full parameter space (mass, misalignment, and bias amplitude) results in a degradation of the achievable upper bound by approximately $0.3\%$. To illustrate the full variability of the result within the prior range for $b_g$, we show in the right panel of fig.~\ref{fig:ptes} the impact as a green shaded region. The smallest values of $b_g$ degrade the bound by up to an order of magnitude, while the largest values improve the constraint by roughly half an order of magnitude.

\bibliographystyle{JHEP}
\bibliography{ref}

\end{document}